\documentclass[14pt,aps,preprint,longbibliography]{revtex4-1}
\usepackage[utf8]{inputenc}
\usepackage[T1]{fontenc}
\usepackage{graphicx}
\usepackage{lineno}
\usepackage{bbold}
\usepackage{color}
\usepackage[normalem]{ulem}
\usepackage{epsfig}
\usepackage{amssymb,amsfonts,amsmath,color}
\usepackage{array}
\usepackage{graphicx} 
\usepackage{float}
\usepackage{ragged2e}
\usepackage[caption=false]{subfig}

\makeatletter
\long\def\@makecaption#1#2{%
  \vskip\abovecaptionskip
  \begingroup
    \small
    \noindent
    \parbox{\linewidth}{%
      \justifying
      #1: #2%
    }%
  \endgroup
  \vskip\belowcaptionskip
}
\makeatother

\begin{document}

\begin{titlepage}
\begin{center}

{\LARGE \bfseries When do correlations reflect biological similarity in ecological dynamics?  \par}

\vspace{2cm}

{\large
Akiva Goldberg and   
Nadav M. Shnerb \par
}

\vspace{1cm}

{\small
Department of Physics, Bar-Ilan University, Ramat Gan 52900, 

}\vspace{1cm}

\vfill

\end{center}

\end{titlepage}

\begin{abstract}
   \noindent The structure of competitive ecological communities is shaped by the strength of interactions between species, which in turn reflects their biological similarity. At the same time, the stochastic forcing that drives abundance fluctuations is itself biologically grounded: species that are more similar may be expected to respond more similarly to environmental variation. This motivates the increasingly common use of correlations in abundance time series, particularly in microbial communities, as proxies for biological similarity or niche overlap. Here we analyze the relation between biological similarity and abundance correlations in stochastic community models. We require that the stochastic forcing acting on different species be correlated in proportion to their biological similarity, and ask how such forcing is reflected in abundance correlations. We show that this requirement cannot, in general, be satisfied within the widely used stochastic Lotka-Volterra framework, and that even when it is, abundance correlations carry no information about niche overlap. In contrast, consumer–resource models provide a natural framework for biologically grounded stochasticity. In this setting, however, the interpretation of abundance correlations depends strongly on the pathway through which noise enters the system: direct forcing of consumers and resource-mediated fluctuations encode different biological quantities. These results have implications both for the modeling of stochastic ecological communities and for understanding what can, and cannot, be inferred from correlations in community time series. 
\end{abstract}
\maketitle

\section{Introduction}

Temporal environmental stochasticity is a central driver of ecological populations and communities~\cite{lande2003stochastic,fung2022effects}. Multispecies communities therefore reflect the combined action of deterministic interactions and stochastic environmental forcing. Accordingly, they have been approached from a wide spectrum of theoretical perspectives, ranging from frameworks that emphasize deterministic structure~\cite{bunin2017ecological,barbier2018generic} to others in which temporal variation is viewed as predominantly stochastic or chaotic~\cite{kalyuzhny2015neutral,danino2018theory,mallmin2024chaotic,arnoulx2024many,grilli2020macroecological}.

A central problem in this context is whether the resulting fluctuations can be used to infer properties of the underlying system. In ecological communities, this question is especially natural because biological similarity between species is often expected to generate both similar interaction patterns and similar responses to environmental variability. This raises the possibility that abundance correlations measured in community time series may serve as proxies for biologically meaningful quantities such as niche overlap. As time-resolved community data become increasingly available, particularly in microbial ecosystems, such inference-based approaches are attracting growing interest~\cite{goyal2022interactions,sireci2023environmental,crocker2025microbial,chen2025inferring}.

This possibility immediately raises a modeling question: how should environmental stochasticity be introduced if one wishes such correlations to retain a biologically meaningful interpretation? In many stochastic descriptions of community dynamics, environmental variability is incorporated by adding stochastic terms directly to Lotka-Volterra equations~\cite{loreau2013biodiversity,van2003front,kalyuzhny2015neutral}. This construction is technically convenient and is therefore often used as a framework in the literature. Yet it leaves open a basic conceptual issue. If the interaction matrix is interpreted in terms of niche overlap~\cite{chesson2013species,spaak2020intuitive}, what should determine the correlation structure of the stochastic forcing? In most applications of stochastic Lotka-Volterra models, this structure is imposed externally: the noise acting on different species is typically assumed either to be uncorrelated, or to have some prescribed level of correlation that is not derived from the  biological quantities that determine the interactions.

In this paper, we address two closely related questions. First, we ask whether environmental stochasticity in community dynamics can be modeled in a biologically grounded manner, such that species that are mechanistically similar at the deterministic level are also subject to similar stochastic forcing. This, in turn, requires the covariance structure of the noise to track the degree of biological similarity between species, as encoded by their niche overlap. As we show below, however, implementing such a stochastic description is far from trivial.

Second, we ask under what conditions species correlations can be given an inferential interpretation, and in particular whether they can be used to recover niche overlap. We show that the answer depends qualitatively on the mechanism by which stochasticity enters the system. More generally, the inferential content of species correlations cannot be assessed independently of the pathway through which environmental fluctuations are transmitted.

This discussion points directly to a limitation of the standard Lotka-Volterra framework. Although LV equations provide an adequate minimal description of deterministic community dynamics~\cite{bunin2017ecological,barbier2018generic}, their stochastic extension is much more constrained. As we show below, introducing environmental noise directly at the LV level generally does not allow for a biologically grounded covariance structure, and even when such a construction is feasible, it eliminates the inferential content of abundance correlations. The standard stochastic LV framework is therefore limited not only as a model of environmental forcing, but also as a basis for inference from community time series.

This limitation is not merely technical, but reflects a fundamental constraint of coarse-grained descriptions: once the mechanistic origin of interactions is suppressed, it becomes impossible to impose a biologically grounded structure on the stochastic forcing. This motivates a shift to a more mechanistic level of description. A natural framework for doing so is provided by consumer–resource models~\cite{sakarchi2025macarthur,sireci2023environmental,chen2025inferring}. Within this framework, it becomes possible to distinguish between noise that acts directly on the growth rates of consumers, independently of resource dynamics (``external'' noise), and noise that acts indirectly through resource availability when the latter fluctuates stochastically in time. We show that these two forms of stochasticity generate qualitatively different patterns of abundance correlations: the former reflects niche overlap, whereas the latter depends instead on a different quantity, the yield-depletion mismatch, which was analyzed in a recent study~\cite{goldberg2025niche}. A short presentation of this quantity is provided in Appendix \ref{app:ydm}. 

In the earlier work mentioned above~\cite{goldberg2025niche}, the central question was interpretive: given a consumer-resource system with stochastic fluctuations at the resource level, what do the resulting correlations reveal about the underlying mechanisms that shape the community. Here we step back and address a more fundamental question: how should stochasticity be modeled in the first place if it is to remain biologically grounded, and how does the choice of modeling level constrain the use of correlations for inference. The present work therefore provides a broader framework linking three issues that are often treated separately: the modeling of environmental stochasticity, the mathematical realizability of a biologically grounded noise structure, and the inferential content of the resulting fluctuations.

The paper is organized as follows. In Sec.~\ref{sec2}, we present the deterministic consumer-resource and Lotka-Volterra descriptions and establish the connection between them. In Sec.~\ref{sec3}, we analyze stochastic Lotka-Volterra dynamics and show that imposing a biologically grounded covariance structure is generically obstructed by a realizability constraint, and that even when such a construction is feasible it yields no informative abundance correlations. In Sec.~\ref{sec4}, we turn to stochastic consumer-resource dynamics and show that a biologically grounded treatment of environmental noise is possible only at that level, where direct and resource-mediated fluctuations give rise to distinct statistical signatures. Finally, we discuss the implications of these results for the modeling of environmental stochasticity and for inference from correlations in community dynamics.

\section{From consumer-resource dynamics to Lotka-Volterra: effective interactions and niche overlap} \label{sec2}

To make the discussion concrete, we consider two standard modeling approaches for community dynamics: the consumer-resource (CR) model and the Lotka-Volterra (LV) model. These should be viewed not as unrelated alternatives, but as two levels of description of the same interacting system. Although we expect our conclusions to hold more broadly, this pair of models provides an especially transparent framework: the CR description retains the mechanistic distinction between resource depletion and consumer benefit, whereas the LV description emerges from it in appropriate limiting regimes as an effective coarse-grained dynamics. This relation will be central below, since the passage from CR to LV preserves niche overlap while potentially obscuring other mechanistic information.

We consider a community of $S$ consumer species, competing for $Q$ resources. The abundance of consumer species $i$ is denoted by $n_i(t)$, and the biomass of resource $k$ by $R_k(t)$. Consumer growth is determined by resource uptake, whereas each resource grows logistically in the absence of consumers and is depleted through consumption. The deterministic dynamics is therefore governed by the coupled equations
\begin{align} \label{eq1}
\frac{dn_i(t)}{dt} &=
- n_i
+ n_i \sum_{k=1}^{Q} \gamma_{ik} R_k, \nonumber
\\
\frac{dR_k(t)}{dt} &=
R_k - R_k^2
- R_k \sum_{i=1}^{S} \lambda_{ki} n_i .
\end{align}
Here $\gamma_{ik}$ is the $(i,k)$ entry of the $S\times Q$ matrix $\Gamma$ and quantifies yield, namely the increase in the growth rate of consumer $i$ per unit of resource $k$. Similarly, $\lambda_{ki}$ is the $(k,i)$ entry of the $Q\times S$ matrix $\Lambda$ and quantifies depletion, namely the per-capita removal of resource $k$ by consumer $i$.

A key feature of the present CR formulation is that it keeps separate two mechanistically distinct quantities: the benefit that a consumer derives from a resource, encoded in $\Gamma$, and the depletion that the same consumer imposes on that resource, encoded in $\Lambda$. In traditional consumer-resource models~\cite{sakarchi2025macarthur}, the yield and depletion coefficients are typically assumed to be proportional at the level of individual consumer-resource interactions, $\gamma_{ik} = c_k \lambda_{ki}$, where the proportionality factor depends only on the resource, thereby identifying consumption and growth up to a fixed conversion efficiency. Here we relax this assumption and allow depletion and benefit to be misaligned. Such a decoupling may arise from differences in conversion efficiency, wasteful or destructive uptake, metabolic by-products, cross-feeding, and related processes~\cite{Postma1989,Bjorge2018,Basan2015,Speakman2005,Vucetich2004,OwenSmith1988,Jansen2004,Gosling2001}.  Consistent with this view, recent studies increasingly suggest that the yield-to-depletion ratio is not fixed, but can vary across conditions, resources, and taxa~\cite{liu2024ecosystem,gibbs2022stability,blumenthal2024phase}. This separation will prove essential below, as it introduces an additional degree of freedom that is invisible at the LV level but directly controls the response to stochastic resource fluctuations. 

To obtain an effective Lotka-Volterra description, one integrates out the resource dynamics and replaces them by direct competition terms between consumer species. This quasi-steady-state reduction assumes that
\[
\frac{dR_k}{dt}=0,
\]
so that resource abundances can be expressed as functions of consumer abundances and substituted back into the consumer equations. This approximation is justified in two closely related limits: when resource dynamics is much faster than consumer dynamics, so that resources may be treated as fast variables, and when the system is considered close to equilibrium. The resulting effective dynamics takes the Lotka-Volterra form
\begin{equation} \label{eq3}
\frac{dn_i}{dt}
=
n_i \left(
1-
n_i
-
 \sum_{j \neq i} \alpha_{ij} n_j \right),
\end{equation}
with effective interaction matrix
\begin{equation} \label{eq_alpha}
\boldsymbol{\alpha}=\Gamma\Lambda.
\end{equation}
This reduction preserves the effective niche-overlap structure, now encoded in $\boldsymbol{\alpha}$, but it eliminates the mechanistic distinction between resource depletion and consumer benefit that remains explicit in the CR description.

The effective matrix $\boldsymbol{\alpha}$ will play a central role throughout this paper. In the competitive setting, when $\alpha_{i,i}$ is normalized to one, $\alpha_{ij}$ is naturally interpreted as a measure of the effective overlap between consumer species $i$ and $j$~\cite{chesson2012species,spaak2020intuitive}. For a non-symmetric interaction matrix, the corresponding quantity may be taken to be the symmetrized combination $(\alpha_{ij}+\alpha_{ji})/2$. When this overlap vanishes, the two consumers effectively exploit distinct resources and do not constrain one another's growth. By contrast, a value close to one corresponds to nearly maximal overlap. For example, if a single biological species were artificially partitioned into two populations according to a marker irrelevant to growth and resource uptake, the interaction coefficient between the two populations would be $\alpha=1$.

\section{Biologically-grounded stochastic dynamics: the failure of the  Lotka-Volterra model} \label{sec3}

\subsection{A biologically grounded noise prescription}

We now turn to the stochastic extension of the effective Lotka-Volterra description. If environmental variability is introduced directly at the LV level, the standard prescription~\cite{loreau2013biodiversity,van2024tiny,kalyuzhny2015neutral} is to add a time-dependent random term to the linear growth rate. This leads to the stochastic Lotka-Volterra equations used in much of the literature,
\begin{equation} \label{eqLV1}
\frac{dn_i}{dt}
=
n_i
-
n_i^2
-
n_i \sum_{j \neq i} \alpha_{ij} n_j
+
\sigma_e \eta_i(t)\, n_i ,
\end{equation}
or, equivalently, in stochastic differential equation form,
\begin{equation}
dn_i
=
\left(
n_i
-
n_i^2
-
n_i \sum_{j \neq i} \alpha_{ij} n_j
\right) dt
+
\sigma_e n_i\, dW_i ,
\end{equation}
where $\langle \eta_i(t) \rangle = 0$ and $\langle \eta_i(t)\eta_i(t') \rangle = \delta(t-t')$, and $W_i$ denotes a Wiener process. In the simulations reported below, we implemented the Stratonovich convention, although the qualitative picture is independent of this choice.  

The key modeling question is then the correlation structure of the stochastic forcing. What should one assume about the correlations between the two timeseries, $\eta_i(t)$ and $\eta_j(t)$? 

In most of the literature, the stochastic terms $\eta_i(t)$ and $\eta_j(t)$ are taken either to be completely independent or to exhibit some arbitrarily prescribed level of correlation. Although this approach is technically straightforward, its ground in biological reality is problematic. 

We suggest that a biologically grounded correlation structure requires the covariance matrix of the noise,
\[
C_{ij} = \langle \eta_i(t)\eta_j(t) \rangle ,
\]
to be, at least to first approximation, proportional to $\alpha_{ij}$, for the following reasons:
\begin{enumerate}
    \item The coefficient $\alpha_{ij}$ reflects the extent to which species depend on similar resources. Accordingly, the correlations between their environmental responses are expected to increase with $\alpha_{ij}$: species with larger niche overlap should experience more similar fluctuations in their growth rates.

    \item Even setting aside resource dynamics, the interaction coefficient $\alpha_{ij}$ often reflects biological similarity between species, for example phylogenetic relatedness. It is therefore natural to expect that external environmental factors, such as temperature or pH, will affect species in a similar manner when $\alpha_{ij}$ is large.
\end{enumerate}

We are therefore led to model stochasticity in the LV dynamics by introducing $S$ stochastic processes $\eta_i(t)$, with $i=1,\dots,S$, whose covariance matrix satisfies
\begin{equation}
C_{ij}
=
\alpha_{ij}.
\end{equation}
This prescription should not be viewed as a unique biological law, but rather as a canonical limiting case of biologically grounded direct noise. More generally, one may expect the covariance structure to reflect biological similarity only approximately. The choice $C^{(\eta)}_{ij} = \alpha_{ij}$ therefore provides a stringent test case: if the stochastic Lotka-Volterra framework fails even under this maximally favorable assumption, it cannot serve as a reliable basis for correlation-based inference.

A set of correlated time series of that kind can be generated, for example, using the Cholesky decomposition~\cite{golub2013matrix}. At first sight, this prescription offers a natural route to biologically grounded stochasticity in the LV dynamics. As we now show, however, it encounters two major obstacles: a realizability problem and an inferential one. 

\subsection{The complexity-stochasticity problem}

For symmetric interaction matrices, the prescription above - that is, using the Cholesky decomposition to generate a biologically grounded correlation structure - is feasible only if the matrix $\boldsymbol{\alpha}$ is positive semidefinite. Thus, biologically grounded stochastic forcing with $C_{ij} = \alpha_{ij}$ can be imposed only when the effective interaction matrix satisfies a very restrictive spectral condition.

This requirement is closely analogous, in spirit, to May's classical stability criterion~\cite{may1972will}. In May's case, coexistence is limited by the requirement that the Jacobian around a stable solution of Eq. (\ref{eq3}) satisfy an appropriate spectral stability condition, whereas here the Cholesky construction imposes a comparable spectral constraint directly on the interaction matrix $\boldsymbol{\alpha}$. The analogy is not one of exact mathematical equivalence, but of statistical structure: in both settings, a large heterogeneous matrix must satisfy the same restrictive spectral condition, and in both settings this condition becomes increasingly unlikely to hold as the matrix grows larger and the variance of its entries increases. If the Jacobian and $\boldsymbol{\alpha}$ are assumed to have comparable statistical properties, then the two criteria should be expected to fail in the same regime of large size and strong heterogeneity.

In practice, this analogy is reflected in the same qualitative pattern. Whenever the LV equations admit stable coexistence of all $S$ species, it is generally possible to perform a Cholesky decomposition and construct stochastic fluctuations whose covariance structure follows the niche-overlap matrix $\boldsymbol{\alpha}$. By contrast, when the deterministic LV system does not admit coexistence, a covariance matrix consistent with this prescription is typically not realizable. In this sense, beyond the classical complexity-diversity problem, one encounters here a related complexity-stochasticity problem: in sufficiently large and heterogeneous communities, biologically grounded stochastic forcing is generically not realizable.

\subsection{Vanishing abundance correlations} 

At first sight, one might hope to avoid the realizability problem discussed above by restricting attention to the subset of species that admits a stable coexistence state. We refer to such a subset as a coexistence clique~\cite{fried2016communities}. Within a coexistence clique, the interaction matrix does, in general, admit a Cholesky decomposition, allowing one to generate stochastic environmental fluctuations with covariance matrix proportional to $\boldsymbol{\alpha}$ and to use these fluctuations in the stochastic LV dynamics.

\begin{figure}
\centering
  \includegraphics[width=12.3cm]{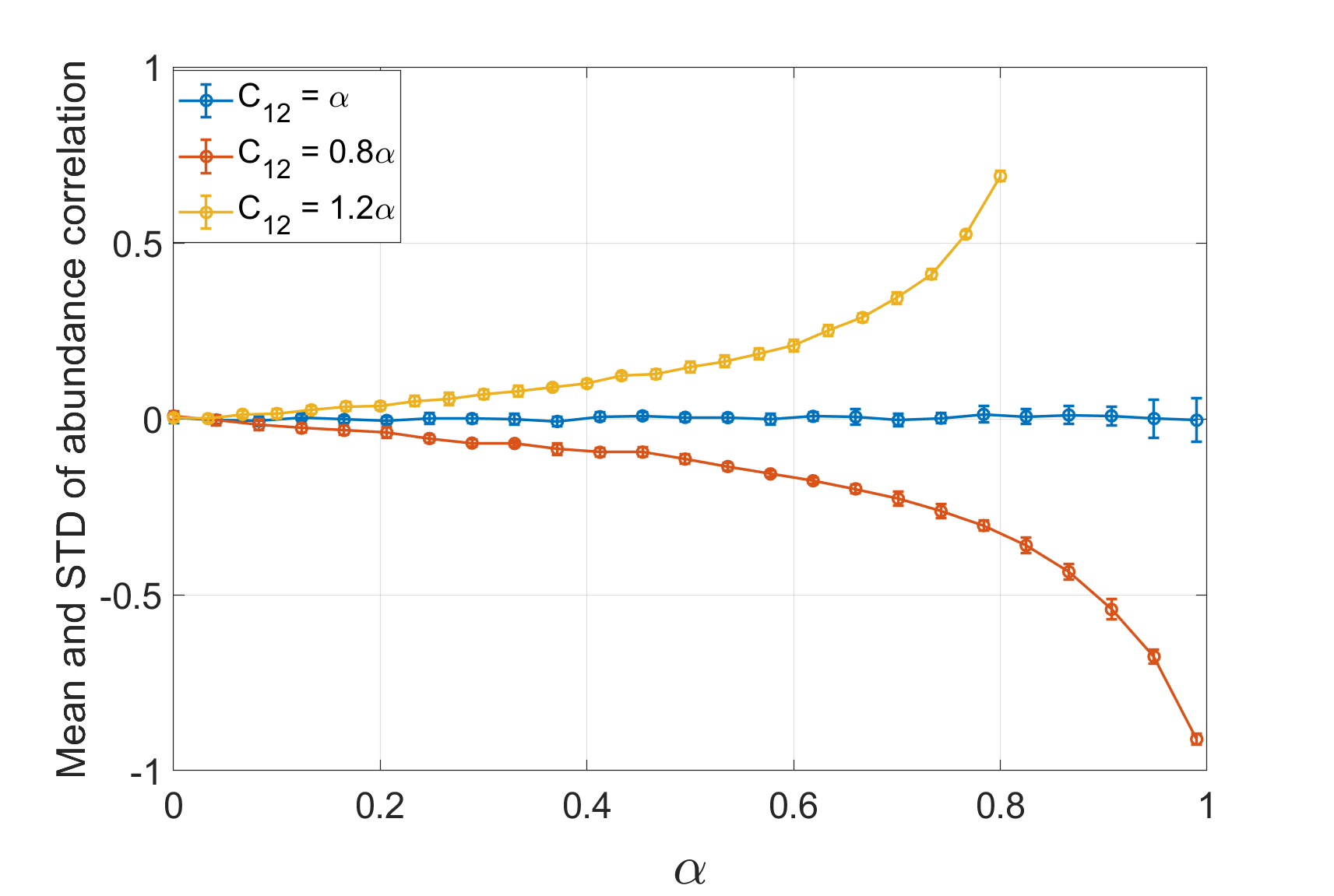}
\caption{{\bf Abundance correlations in the Lotka-Volterra model for two-species community with correlated noise.} The stochastic Lotka-Volterra model (Eqs. \ref{eqLV1}) was simulated for two species, $1$ and $2$. The competition is symmetric, $\alpha_{1,2}=\alpha_{2,1} \equiv \alpha$. The covariance between the corresponding Wiener processes is proportional to $\alpha$, with $C_{1,2} = \langle \eta_1(t)\eta_2(t) \rangle = b_1  \alpha$, where the constant $b_1$ takes the values $0.8, 1$ and $1.2$.  While higher niche overlap $\alpha$ yields negative correlations due to competition, the higher covariance between noise timeseries reflect shared response to environmental variations, hence yield positive correlations. At the point of biologically-grounded relationships, $b_1 = 1$, the two effects cancel each other. Positive correlations appear when the shared response is dominant ($b_1 >1$) and correlations become negative when competition is dominant. Results were extracted from simulations of $T=100,000$ timesteps with $\sigma_e = 0.05$, where the mean and the variance were calculated using the last $9$ time windows of $10,000$ timesteps each.       }
\label{fig1}
\end{figure}

However, even within a coexistence clique, a second and more fundamental difficulty remains. If the noise covariance is chosen to be proportional to $\alpha_{ij}$, abundance correlations vanish in a two-species community. This result is derived in the weak-noise limit, by linearizing the dynamics around a stable coexistence fixed point (see Appendix \ref{app:Corr_calc}).  The cancellation  is illustrated numerically in Fig.~\ref{fig1}. The reason is that, under this prescription, the same coefficient $\alpha_{ij}$ controls two opposing effects: it determines how similarly species respond to environmental fluctuations and, at the same time, how strongly they compete with one another. Thus, whenever two species are driven in the same direction by the environment, they also inhibit one another more strongly. Within the stochastic LV structure, the positive correlation induced by the shared environment is therefore exactly canceled by the negative correlation induced by competition.

The implication for inference is immediate. Within this stochastic LV framework, abundance correlations carry no information about niche overlap. Even under the most favorable noise prescription, namely, one in which the covariance of the stochastic forcing is chosen to follow $\boldsymbol{\alpha}$, abundance-level inference fails. The same result holds (see Appendix \ref{app:Corr_calc}) in a larger community with many species, provided that the interaction matrix $\boldsymbol{\alpha}$ is symmetric.

\section{Biologically grounded stochastic dynamics in the consumer-resource model} \label{sec4}

We now return to the consumer-resource level, precisely because it preserves the mechanistic distinction that is lost in the coarse-grained Lotka-Volterra description. In particular, it allows us to distinguish between stochasticity that acts directly on consumers and stochasticity that acts indirectly through resources. We therefore consider the stochastic version of the consumer-resource model, Eq.~(\ref{eq1}),
\begin{align} \label{eq10}
\frac{dn_i(t)}{dt} &=
- n_i
+ n_i \sum_{k=1}^{Q} \gamma_{ik} R_k
+ \sigma_e^c \xi_i(t)\, n_i, \nonumber
\\
\frac{dR_k(t)}{dt} &=
R_k - R_k^2
- R_k \sum_{i=1}^{S} \lambda_{ki} n_i
+ \sigma_e^r \zeta_k(t)\, R_k.
\end{align}
Here, both $\xi_i(t)$ and $\zeta_k(t)$ are white-noise processes. The terms proportional to $\sigma_e^r$ describe stochasticity in the growth rates of the resources, which in turn affects consumers indirectly through resource availability. The terms proportional to $\sigma_e^c$, by contrast, represent the direct effect of external environmental factors, such as temperature or pH, on consumer dynamics.

We assume that stochasticity at the resource level ($\sigma_e^r$ terms) is uncorrelated across different resource types. The underlying idea is that interaction with consumers is not, in many cases, a biologically meaningful evidence or source of similarity among resources. Consumers may affect resource abundances only weakly, and similar patterns of interaction with consumers do not necessarily imply similar physiology among resources. This assumption may admit exceptions, for example in systems where consumers and resources have undergone substantial coevolution, but we do not pursue such cases here. By contrast, we assume, as before, that stochasticity at the consumer level reflects biological similarity and  is thus proportional to the interaction matrix $\boldsymbol{\alpha}$, so $
C_{ij} \equiv \langle \xi_i(t)\xi_j(t) \rangle = \alpha_{i,j}$.

We can now evaluate these correlations, both numerically and analytically, within a biologically grounded framework. This analysis yields two important observations. 

First, let us consider the case $\sigma^r_e = 0$, i.e., when stochasticity affects only the consumer species via external mechanisms (such as temperature or pH) and not through the resources.  This reveals a major difference between Lotka-Volterra dynamics and the consumer-resource system. In the LV case, when the noise is biologically grounded, i.e., $C_{ij}=\alpha_{ij}$, 
the effect of competition, encoded by $\alpha$ and tending to induce negative correlations, exactly cancels the effect of the shared environmental response, which tends to induce positive correlations. The resulting abundance correlation is therefore zero.

\begin{figure}
\centering
\begin{minipage}{0.48\linewidth}
\centering
\includegraphics[width=\linewidth,height=0.32\textheight,keepaspectratio]{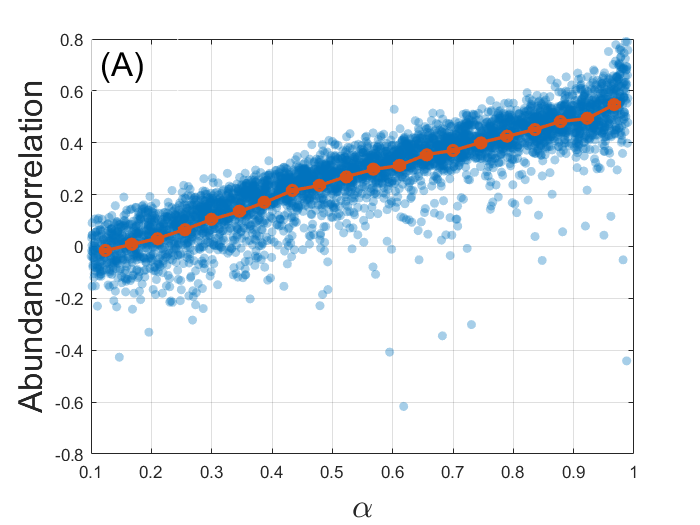}

\end{minipage}
\hfill
\begin{minipage}{0.48\linewidth}
\centering
\includegraphics[width=\linewidth,height=0.32\textheight,keepaspectratio]{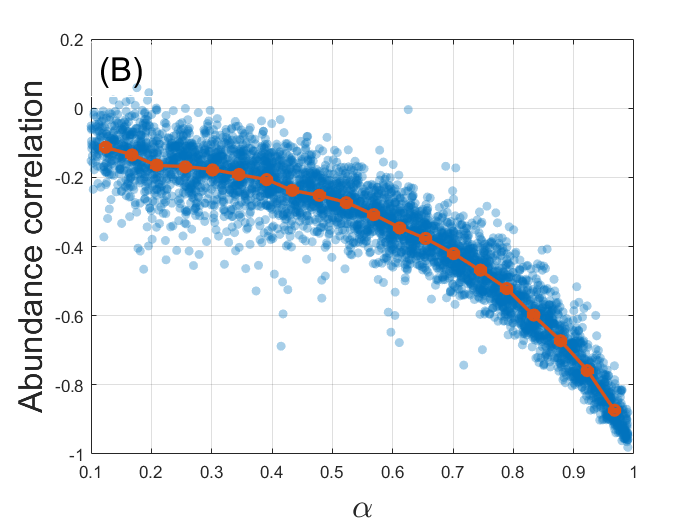}

\end{minipage}

\caption{{\bf Abundance correlations in consumer-resource system with consumer noise.} The stochastic model (Eqs. \ref{eq10}) were simulated for two species and three resources. Niche overlap $\alpha$ is related to the matrices $\Gamma$ and $\Lambda$ through Eq. (\ref{eq_alpha}). Shown here are the abundance correlations between consumer species, as obtained from long simulations, as a function of $\alpha$.  Since the mapping is highly redundant, many combinations of $\Gamma$ and $\Lambda$ correspond to the same value of $\boldsymbol{\alpha}$~\cite{goldberg2025niche}, which brings some "width" to the resulting outcome. In Panel A (left), the noise is biologically grounded, with
\(C_{12}=\langle \xi_1(t)\xi_2(t')\rangle=\alpha\delta(t-t')\).
In this case, abundance correlations are positive and increase with \(\alpha\),
because the shared response to environmental fluctuations dominates the negative
effect of competition. In Panel B (right), the stochastic forcing is uncorrelated,
\(C_{12}=0\). The shared environmental response is then absent, so competition
dominates: abundance correlations are generally negative and become more negative
as \(\alpha\) increases.}
\label{fig2}
\end{figure}

In the present case, however, the same value of $\alpha$ is obtained through Eq. (\ref{eq_alpha}), under the mapping to Lotka-Volterra, and hence the same degree of niche overlap. Yet, the qualitative outcome is different. The reason is that competition now acts more slowly, because it is mediated through resource dynamics, whereas the shared environmental response is more immediate and therefore has a stronger effect. As a consequence, biologically grounded stochasticity produces a positive abundance correlation. Negative correlations arise only when the shared environmental component is reduced, as in the case of stochastic forcing that is not structured according to $\alpha$, allowing the effect of competition to dominate. This behavior is illustrated in Fig.~\ref{fig2}.

Since this deviation from the LV behavior is directly related to the slower response of resource-mediated interactions, it disappears when  resources become fast variables and adapt to the instantaneous concentration of the consumers.  In that limit the CR system is expected to approach the effective LV description more closely. Consistent with this expectation, our simulations suggest (results not shown) that, in the biologically grounded case $C_{ij}=\alpha_{ij}$, when the resource dynamics is much faster than that of the consumer, the abundance correlations indeed tend toward zero, as in the LV model. It should be noted, however, that this fast-variable limit is difficult to realize when $\alpha \to 1$. In that regime, consumer and resource modes become strongly coupled, and the relaxation timescale to equilibrium diverges for both.  

Our second observation is that a  qualitatively different behavior arises for resource-mediated fluctuations.  If the only source of fluctuations is the resource ($\sigma_e^c=0$),  the consumer-consumer correlation function is  almost independent of $\alpha_{ij}$. As discussed  in our recent work~\cite{goldberg2025niche}, correlations in that case depend on the yield-depletion mismatch YDM, that reflects the imbalance between competition and shared response, see Appendix \ref{app:ydm}.

To see this intuitively, consider two species, $i$ and $j$, that deplete resources similarly, but for which one of them, say species $i$, derives a much greater benefit from a given resource. An increase in the availability of that resource will then preferentially boost species $i$, whose growth will in turn deplete the relevant resources more strongly,  inducing a decrease in the abundance of $j$ and thus negative correlation between the two species. Conversely, if depletion patterns differ while yields remain similar, the resulting correlations are positive. Resource-mediated correlations therefore reflect the yield-depletion mismatch $YDM$, rather than the niche overlap $\alpha$.

\begin{figure} 
\centering

\begin{minipage}{0.48\linewidth}
    \centering
    \includegraphics[width=\linewidth,height=0.32\textheight,keepaspectratio]{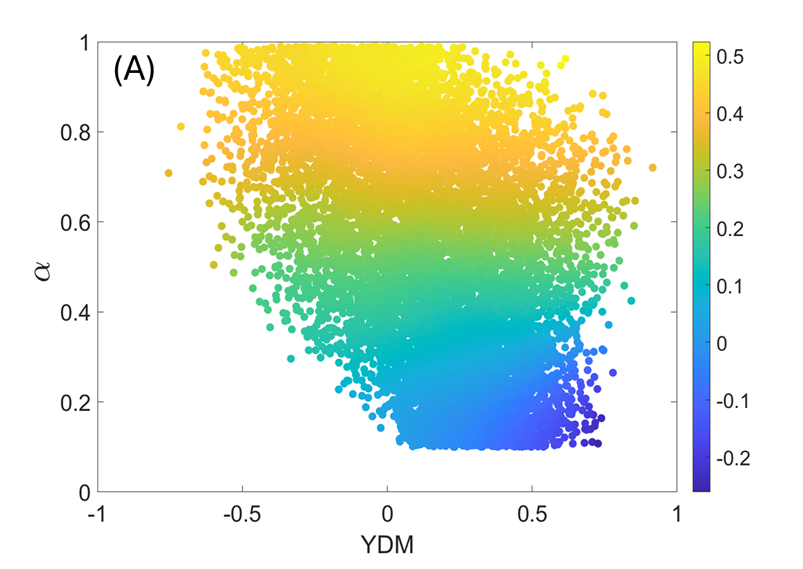}
\end{minipage}
\hfill
\begin{minipage}{0.48\linewidth}
    \centering
    \includegraphics[width=\linewidth,height=0.32\textheight,keepaspectratio]{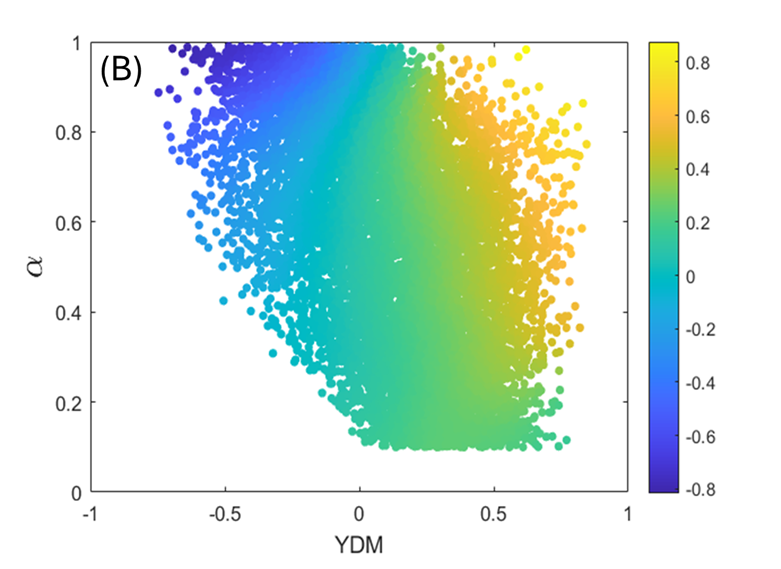}
\end{minipage}

\caption{{\bf Direct and resource-mediated noise.} 
In both panels, color (see colorbar) indicates the level of correlation obtained from numerical experiments across the $YDM$-$\alpha$ plane. In the left panel the stochasticity is external ($\sigma_e^r=0$, no noise term in the resource equation) and in the right panel there is no external noise ($\sigma_e^c=0$). Clearly, external noise yields correlations that are a function of $\alpha$, not of $YDM$, while in resource-mediated stochasticity the role of niche overlap $\alpha$ is negligible and $YDM$ governs the correlations. We implemented a two-consumer, three-resource model, and we generate many randomly chosen $\Gamma$ and $\Lambda$ where $\alpha_{1,2} = \sum_j \Gamma_{1,j} \Lambda_{j,2}$, see Supplementary \ref{app:gammalambda}.}
\label{fig3} 
\end{figure}

This result, and the sharp distinction between resource-mediated stochasticity and external stochasticity, are demonstrated in  Fig.~\ref{fig3}. The value of abundance correlations is $YDM$-dependent for $\sigma_e^c=0$ and is $\alpha$-dependent when $\sigma_e^r=0$. 

The consumer-resource framework therefore resolves the ambiguity that arises at the Lotka-Volterra level by keeping distinct sources of stochasticity mechanistically separate. Direct consumer-level fluctuations carry information about niche overlap, whereas resource-mediated fluctuations carry information about yield-depletion mismatch. As a result, abundance correlations are not intrinsically informative: their interpretation depends on the origin of the noise. Any attempt to use such correlations for inference therefore requires an understanding how stochasticity enters the system.

We would like to note a  related work by \citet{chen2025inferring}, who also studied the inference of resource-mediated interactions from time-series data within a consumer-resource framework. A central conclusion of their work is consistent with ours: equal-time abundance correlations are generally unreliable indicators of resource overlap. One mechanism underlying this failure in their case is that changes in the timescale of environmental forcing shift the balance between competition and shared response.  These authors addressed the limitations of zero-time correlations by developing dynamical observables, in particular cross-power spectral density and coherence, which retain lag- and frequency-dependent information and can recover resource guild structure. We do not pursue such dynamical observables here, as our focus is not on improving inference from abundance time series, but rather on identifying the mechanistic conditions under which abundance correlations admit a meaningful interpretation. 

Despite these similarities, there are several important differences in the modeling assumptions underlying our approach and that of \citet{chen2025inferring}. First, in their construction the effective interaction matrix is not normalized, so that diagonal self-interaction terms vary across species and may mix niche overlap with total resource use. Second, their parametrization limits differences between yield and depletion, corresponding in our terminology to a regime of small yield-depletion mismatch. Third, environmental fluctuations in their model are temporally correlated, either periodic or Ornstein-Uhlenbeck, and enter exclusively through the resources. In contrast, we employ white-noise forcing and explicitly distinguish between direct consumer-level noise and resource-mediated fluctuations.

\section{Discussion}

The first conclusion of this work is that biologically grounded stochasticity is generically incompatible with large, heterogeneous communities. If niche overlap is interpreted as biological similarity, it is natural to require the covariance of the stochastic forcing to track the interaction matrix $\boldsymbol{\alpha}$. However, this is possible only when $\boldsymbol{\alpha}$ satisfies the spectral constraints of a covariance matrix~\cite{golub2013matrix}. 

This leads to a direct analogue of May's complexity-diversity argument~\cite{may1972will}. Just as increasing diversity and heterogeneity make stable coexistence unlikely, they also make it unlikely that stochastic fluctuations can consistently reflect biological similarity. This defines a correlation-diversity problem: increasing diversity does not only destabilize deterministic coexistence, but also limits the ability of stochastic fluctuations to encode biological similarity.

This connection can be understood intuitively by considering the relation between the two limitations discussed above. When a community violates May's complexity-diversity criterion, many species cannot persist deterministically and survive only marginally, for example through continuous immigration. In that regime, the fluctuations experienced by these marginal species are dominated by external inputs and by the collective dynamics of the persistent community, rather than by their pairwise biological similarity. One should therefore not expect the stochastic covariance structure to track niche overlap outside the coexistence clique. 

A second conclusion is that biologically grounded stochasticity cannot be modeled directly at the level of Lotka-Volterra equations. Even in the low-diversity cases where the covariance prescription $C_{ij}\propto \alpha_{ij}$ is realizable, the resulting abundance correlations vanish identically. The reason is that, within the stochastic LV description, the same coefficient $\alpha_{ij}$ reflects two opposing trends: the biological similarity between species and between the environmental forcing acting on them, and the strength of the competition between them. As a consequence, the positive correlation induced by a shared environmental response is exactly canceled by the negative correlation induced by stronger competition. The LV formalism therefore builds in a hidden symmetry between gain and loss, or equivalently between shared response and competitive feedback, and this symmetry destroys the abundance-level signal. For this reason, inference based on abundance correlations cannot rely on stochastic LV modeling: the question of what abundance correlations 
reveal about niche overlap has, within that framework, no informative answer. 

The situation changes qualitatively in the consumer-resource dynamics. There, biologically-grounded stochasticity can be introduced at the mechanistic level, and one must distinguish between direct noise acting on consumers and indirect noise transmitted through resources. These two sources of stochasticity have fundamentally different signatures. Direct and biologically grounded consumer-level noise generates correlations that track $\alpha_{ij}$ and therefore retain information about niche overlap or, more generally, about mechanistic similarity between species (guild structure). By contrast, resource-mediated noise does not reflect $\alpha_{ij}$; instead, as shown in our previous work~\cite{goldberg2025niche}, it reflects the yield-depletion mismatch (YDM), namely the extent to which depletion and benefit are misaligned between species. In realistic systems, one should therefore expect a mixture of these two contributions. This immediately implies that abundance correlations can be used for inference only if one has some independent understanding of the origin of the noise and of the pathway through which it enters the system. 

As mentioned above, the recent work by \citet{chen2025inferring} showed that abundance time series may nevertheless contain recoverable information about resource competition when analyzed dynamically rather than through equal-time correlations: cross-power spectral density and coherence can reveal resource guilds in consumer-resource systems with time-dependent resource supply. Our results are complementary. They show that the interpretation of abundance-derived statistics depends on the mechanism of environmental forcing: direct consumer-level fluctuations, resource-mediated fluctuations, and coarse-grained LV noise need not encode the same biological quantity.


Finally, several caveats should be kept in mind. First, the assumption that stochastic forcing at the consumer level should be proportional to $\boldsymbol{\alpha}$ is biologically natural, but not universal. Species may converge to similar niche positions while retaining very different physiologies (e.g., through convergent evolution), in which case their responses to external variables such as temperature or pH need not be similar. Second, we assumed here that resource-level stochasticity is uncorrelated across resource types. However, one could equally imagine systems in which resources that interact similarly with the consumers respond in correlated ways, suggesting covariance structures for the resource noise that are tied to $\Gamma$ or $\Lambda$. Such possibilities are especially plausible in systems with strong coevolution or tightly structured resource classes. Extending the present framework to these cases is an important direction for future work.

{\bf Acknowledgments:} N.M.S acknowledges support from the Israel Ministry of Science (Italy-Israel cooperation, grant no. 7578) and of the Israel Science Foundation (grant no. 2435/24).

\bibliography{ref}

\newpage
\appendix

\begin{center}

\section{The yield-depletion mismatch (YDM)}
\label{app:ydm}
\end{center}

In the main text, we emphasized that when stochasticity is transmitted through resource fluctuations, abundance correlations between consumers are not primarily governed by niche overlap alone. Instead, they depend on an additional mechanistic property of the consumer-resource system, which termed, in a previous work, the \emph{yield-depletion mismatch} (YDM) \cite{goldberg2025niche}. In this appendix, we briefly introduce the yield-depletion mismatch (YDM) and outline its biological interpretation.

\subsection{Resource-mediated correlations are controlled by yield-depletion mismatch}

Consider two consumer species \(i\) and \(j\) relying on a common pool of resources. At the level of the reduced Lotka-Volterra description, their level of competition is encoded by the effective interaction coefficient $\alpha_{ij}$. In a consumer-resource system $\alpha_{ij}$ captures their niche overlap between these two species.

However, the same value of $\alpha_{ij}$ can arise from many different combinations of consumer traits. In particular, as explained in the main text, each consumer species is characterized by two distinct profiles:

\begin{itemize}
    \item a \emph{yield profile}, describing how strongly each resource contributes to its growth;
    \item a \emph{a depletion profile}, describing the per-capita rate at which the consumer removes each resource from the environment
\end{itemize}

These two profiles need not coincide. Two species may deplete resources in very similar ways while differing strongly in the benefit they derive from them or vice versa. As a result, two consumer-resource systems with the same niche overlap may nevertheless respond very differently to stochastic resource fluctuations.

\subsection{Definition}

We distinguish between two aspects of functional dissimilarity, namely yield dissimilarity $\Delta_\gamma$ and depletion dissimilarity $\Delta_\lambda$, defined through the cosine distance between the relevant rows of the yield matrix $\Gamma$, and columns of the depletion matrix $\Lambda$ (see Eq. (\ref{eq1}) of the main text):
\[
\begin{array}{c}
\Delta_\gamma = 1 - \cos \theta_{i,j}(\gamma) \equiv 1 - \dfrac{\vec{\gamma}_i \cdot \vec{\gamma}_j}{|\vec{\gamma}_i||\vec{\gamma}_j|} \\[6pt]
\Delta_\lambda = 1 - \cos \theta_{i,j}(\lambda) \equiv 1 - \dfrac{\vec{\lambda}_i \cdot \vec{\lambda}_j}{|\vec{\lambda}_i||\vec{\lambda}_j|}.
\end{array}
\]

In what follows, we focus on the \emph{yield-depletion mismatch} $D$, defined by the difference between these two quantities:
\[
D \equiv \Delta_\lambda - \Delta_\gamma.
\]

Thus, \(D\) is not another measure of niche overlap. Rather, it measures the imbalance between similarity in yield and similarity in depletion among systems that may have the same effective interaction coefficient \(\alpha_{ij}\).

\subsection{Interpretation}

The sign of $D$ has a simple qualitative meaning.

\begin{figure}[H]
\centering{\includegraphics[width=10cm]{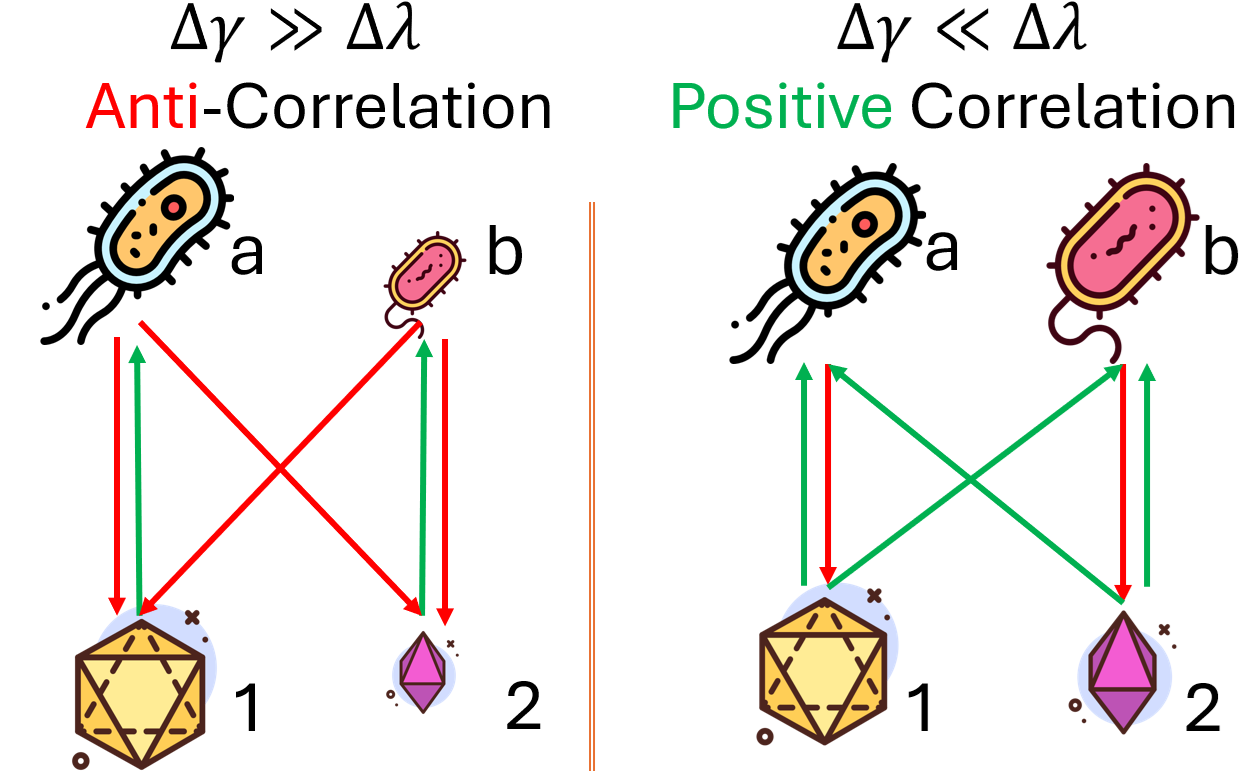}}
\caption{{\bf Intuition Behind the Yield-Depletion Mismatch $D$.}
Two schematic examples corresponding to limiting cases.  Green lines represent yield; red lines represent depletion. 
{\bf Left:} $\Delta\gamma \gg \Delta\lambda$, i.e., each consumer depends on a different resource (green arrows) but depletes (red arrows) both in a similar way.  An increase in one resource strongly benefits only one species; since depletion is not species-specific, this harms the other species, producing negative correlations.  {\bf Right:} $\Delta\gamma \ll \Delta\lambda$. Now, both consumers are generalists and respond similarly to increases in any resource.  Although their depletion profiles differ, the shared response to environmental variation dominates, producing positive correlations. These panels are intentionally stylized and are not meant to represent realistic communities. Natural systems are unlikely to exhibit perfectly generalist or perfectly specialist extremes; instead, yield-depletion asymmetries may arise from species-specific processing, wasteful or destructive uptake, cross-feeding and so on. The panels aim to clarify the mechanisms by which such asymmetries generate positive or negative abundance correlations. }
\label{fig:YDM_intuition}
\end{figure}

When $\Delta_\gamma > \Delta_\lambda$, the two species differ more in yield than in depletion, so $D<0$. In that case, a fluctuation in resource availability may benefit one species much more than the other, while both species still impose similar pressure on the resource pool. This tends to generate negative abundance correlations (see Figure \ref{fig:YDM_intuition}, taken from \cite{goldberg2025niche}).

By contrast, when $\Delta_\lambda > \Delta_\gamma$, one has $D>0$. In that regime, the two species respond more similarly to resource increases, but differ more in the way they deplete resources. As a result, the shared positive response to resource fluctuations tends to dominate, leading to positive abundance correlations.

In other words, YDM quantifies the balance between two competing tendencies:
\begin{enumerate}
    \item \emph{shared response}, which tends to produce positive correlations, and
    \item \emph{effective competition through resource depletion}, which tends to produce negative correlations.
\end{enumerate}

Therefore, abundance correlations due to resource-mediated noise are controlled by $D$ rather than by niche overlap $\alpha$.

\subsection{Relation to previous work}

The role of YDM in the present paper is conceptually distinct from the role it played in \cite{goldberg2025niche}. There, the main question was whether abundance correlations can be predicted from niche overlap in a stochastic consumer-resource system, and the answer was negative: the relevant predictor was instead the yield-depletion mismatch.

Here, our broader goal is to understand when stochasticity can be modeled in a biologically grounded way, and what information can be extracted from the resulting correlations. From that perspective, YDM appears specifically in the case where stochasticity acts \emph{through the resources}. In that case, abundance correlations depend on the mismatch between benefit and depletion, rather than on the effective overlap parameter $\alpha$.

By contrast, when stochasticity acts directly on consumers in a biologically grounded way, correlations track niche overlap much more directly. The key point is therefore that the inferential content of correlations depends not only on the deterministic interaction structure, but also on the mechanistic pathway through which stochasticity enters the system.

\clearpage

\newpage

\begin{center}

    \section{Generating yield and depletion matrices} \label{app:gammalambda}
  \end{center}

\noindent Throughout this paper, we fixed the parameter values in the LV model (thereby also determining the niche overlap) and then searched for many consumer-resource systems that, in the limit ${\dot R}_k = 0$, yield the same $\alpha$ values, using  the redundancy of the mapping from the CR system to LV~\cite{goldberg2025niche}.

To this end, we explored \(\Lambda\) through null-space perturbations:
\begin{itemize}

\item Each of the \(S\) rows of \(\Gamma\) was generated by drawing a random vector of length \(Q\) with positive entries and normalizing it so that the row sum equals 2.

\item Given such a \(\Gamma\), we constructed a compatible \(\Lambda\) by solving:
\[
\Gamma \Lambda = \alpha
\]
\item Since \(\Gamma\) is a \(S \times Q\) matrix, the solution for \(\Lambda\) is underdetermined. The Moore-Penrose pseudoinverse \(\Gamma^\dagger\) provides us with a particular solution:
\[
\Lambda^0 = \Gamma^\dagger \alpha.
\]

\item To explore the solution space, we added to this particular solution a random component from the null space of \(\Gamma\), yielding:
\[
\Lambda = \Lambda^0 + \epsilon \  \text{Null}(\Gamma)
\]
where \(\epsilon\) is a random matrix of appropriate dimensions. 
\end{itemize}
This method ensures that all generated \(\Lambda\) matrices satisfy the required constraints while allowing for variability due to redundancy in the CR-to-LV projection.

\newpage

\begin{center}
\section{Vanishing abundance correlations in the symmetric stochastic Lotka-Volterra model}
\label{app:Corr_calc}
\end{center}
In this appendix we show that, for a symmetric \(S\)-species stochastic Lotka-Volterra system, abundance correlations vanish (to leading order in the weak-noise expansion) when the covariance matrix of the environmental noise is chosen to be equal to the interaction matrix. This calculation generalizes the two-species result and shows that the cancellation is not a special property of the two-dimensional case.

We consider the stochastic Lotka-Volterra system
\begin{equation}
\frac{dn_i}{dt}
=
n_i
\left(
1-\sum_{j=1}^{S}\alpha_{ij}n_j
\right)
+
\sigma_e n_i \eta_i(t),
\qquad i=1,\ldots,S .
\label{eq:stochastic_LV_S_species}
\end{equation}
The matrix \(\alpha\) is assumed to be symmetric,
\begin{equation}
\alpha_{ij}=\alpha_{ji},
\end{equation}
with normalized diagonal entries,
\begin{equation}
\alpha_{ii}=1.
\end{equation}
The environmental noise terms satisfy
\begin{equation}
\left\langle \eta_i(t) \right\rangle =0,
\end{equation}
and
\begin{equation}
\left\langle \eta_i(t)\eta_j(t') \right\rangle
=
C^{(\eta)}_{ij}\delta(t-t').
\end{equation}
Here we focus on the biologically grounded prescription
\begin{equation}
C^{(\eta)}_{ij}=\alpha_{ij}.
\label{eq:noise_cov_equals_alpha}
\end{equation}
Thus, species with larger niche overlap are also assumed to experience more strongly correlated environmental fluctuations.

The deterministic part of Eq.~\eqref{eq:stochastic_LV_S_species} is
\begin{equation}
\frac{dn_i}{dt}
=
n_i
\left(
1-\sum_{j=1}^{S}\alpha_{ij}n_j
\right).
\end{equation}
A coexistence fixed point \(\bar{\mathbf n}\) satisfies
\begin{equation}
1-\sum_{j=1}^{S}\alpha_{ij}\bar n_j=0
\qquad
\text{for all } i,
\end{equation}
or, in matrix form,
\begin{equation}
\alpha \bar{\mathbf n}=\mathbf 1.
\label{eq:fixed_point_condition}
\end{equation}
We assume that this fixed point exists, is positive, and is linearly stable.

We now write the abundance of each species as a small fluctuation around the coexistence fixed point,
\begin{equation}
n_i(t)=\bar n_i+\delta n_i(t).
\end{equation}
To leading order in the fluctuations, the deterministic dynamics becomes
\begin{equation}
\frac{d}{dt}\delta n_i
=
\sum_{j=1}^{S} A_{ij}\delta n_j,
\end{equation}
where \(A\) is the Jacobian matrix evaluated at the coexistence fixed point. From
\begin{equation}
f_i(\mathbf n)
=
n_i
\left(
1-\sum_{k=1}^{S}\alpha_{ik}n_k
\right),
\end{equation}
we obtain
\begin{equation}
A_{ij}
=
\left.
\frac{\partial f_i}{\partial n_j}
\right|_{\mathbf n=\bar{\mathbf n}} .
\end{equation}
Differentiating gives
\begin{equation}
\frac{\partial f_i}{\partial n_j}
=
\delta_{ij}
\left(
1-\sum_{k=1}^{S}\alpha_{ik}n_k
\right)
-
n_i\alpha_{ij}.
\end{equation}
At the fixed point, the first term vanishes because of Eq.~\eqref{eq:fixed_point_condition}. Therefore,
\begin{equation}
A_{ij}
=
-\bar n_i \alpha_{ij}.
\label{eq:jacobian_components}
\end{equation}
It is useful to define the diagonal matrix
\begin{equation}
N
=
\operatorname{diag}(\bar n_1,\ldots,\bar n_S).
\end{equation}
Then Eq.~\eqref{eq:jacobian_components} can be written compactly as
\begin{equation}
A=-N\alpha.
\label{eq:jacobian_matrix_form}
\end{equation}

The stochastic term in Eq.~\eqref{eq:stochastic_LV_S_species} is multiplicative. However, in the weak-noise expansion around the fixed point, its leading-order contribution is obtained by evaluating the amplitude \(n_i\) at \(\bar n_i\). Thus,
\begin{equation}
\sigma_e n_i \eta_i(t)
=
\sigma_e \bar n_i \eta_i(t)
+
O(\sigma_e \delta n_i \eta_i).
\end{equation}
The term \(O(\sigma_e \delta n_i \eta_i)\) contributes only at higher order in the weak-noise covariance calculation. To leading order, the linearized stochastic dynamics is therefore
\begin{equation}
\frac{d}{dt}\delta \mathbf n
=
A\delta \mathbf n
+
B\boldsymbol{\eta}(t),
\label{eq:linearized_stochastic_dynamics}
\end{equation}
where
\begin{equation}
B=\sigma_e N.
\end{equation}
The noise covariance entering the linearized dynamics is therefore
\begin{equation}
D
=
B C^{(\eta)} B^T.
\end{equation}
Using \(C^{(\eta)}=\alpha\), we obtain
\begin{equation}
D
=
\sigma_e^2 N\alpha N.
\label{eq:diffusion_matrix_general}
\end{equation}

Let \(C\) denote the stationary covariance matrix of the abundance fluctuations,
\begin{equation}
C
=
\left\langle
\delta \mathbf n \delta \mathbf n^T
\right\rangle,
\end{equation}
with entries
\begin{equation}
C_{ij}
=
\left\langle
\delta n_i \delta n_j
\right\rangle .
\end{equation}
The linearized white-noise dynamics in Eq.~\eqref{eq:linearized_stochastic_dynamics} implies that the stationary covariance satisfies the Lyapunov equation
\begin{equation}
A C + C A^T + D =0.
\label{eq:lyapunov_general}
\end{equation}
Equation~\eqref{eq:lyapunov_general} follows from the linearized stochastic dynamics and gives the stationary covariance balance for abundance fluctuations around the stable coexistence fixed point.

We now show that Eq.~\eqref{eq:lyapunov_general} has a diagonal solution. Consider
\begin{equation}
C
=
\frac{\sigma_e^2}{2}N.
\label{eq:general_covariance_solution}
\end{equation}
Since \(N\) is diagonal, this proposed covariance matrix has entries
\begin{equation}
C_{ij}
=
\frac{\sigma_e^2}{2}\bar n_i \delta_{ij}.
\end{equation}
In particular, all off-diagonal covariances are zero.

We now verify that Eq.~\eqref{eq:general_covariance_solution} satisfies the Lyapunov equation. Using \(A=-N\alpha\), we have
\begin{equation}
A C
=
(-N\alpha)
\left(
\frac{\sigma_e^2}{2}N
\right)
=
-\frac{\sigma_e^2}{2}N\alpha N.
\end{equation}
Because \(\alpha\) is symmetric and \(N\) is diagonal,
\begin{equation}
A^T
=
(-N\alpha)^T
=
-\alpha N.
\end{equation}
Therefore,
\begin{equation}
C A^T
=
\left(
\frac{\sigma_e^2}{2}N
\right)
(-\alpha N)
=
-\frac{\sigma_e^2}{2}N\alpha N.
\end{equation}
Adding the two terms gives
\begin{equation}
A C + C A^T
=
-\sigma_e^2 N\alpha N.
\end{equation}
Using Eq.~\eqref{eq:diffusion_matrix_general},
\begin{equation}
D
=
\sigma_e^2 N\alpha N.
\end{equation}
Hence,
\begin{equation}
A C + C A^T + D
=
-\sigma_e^2 N\alpha N
+
\sigma_e^2 N\alpha N
=
0.
\end{equation}
Thus, the stationary covariance matrix of the linearized abundance fluctuations is
\begin{equation}
C
=
\frac{\sigma_e^2}{2}
\operatorname{diag}(\bar n_1,\ldots,\bar n_S).
\label{eq:final_covariance_matrix}
\end{equation}

It follows immediately that, for \(i\neq j\),
\begin{equation}
\left\langle \delta n_i \delta n_j \right\rangle
=
C_{ij}
=
0.
\end{equation}
The abundance correlation coefficient between species \(i\) and \(j\) is
\begin{equation}
\rho_{ij}
=
\frac{C_{ij}}{\sqrt{C_{ii}C_{jj}}}.
\end{equation}
Since \(C_{ij}=0\) for all \(i\neq j\), we obtain
\begin{equation}
\rho_{ij}=0,
\qquad i\neq j.
\label{eq:general_zero_correlation_result}
\end{equation}

Therefore, in the symmetric stochastic Lotka-Volterra model, if the covariance structure of the environmental forcing is chosen to match the niche-overlap matrix,
\begin{equation}
C^{(\eta)}=\alpha,
\end{equation}
then all pairwise abundance correlations vanish at leading order in the weak-noise expansion around the coexistence fixed point.

This result shows that the cancellation is a general property of the symmetric
\(S\)-species stochastic Lotka-Volterra model. The same matrix \(\alpha\) controls two
opposing effects. On the one hand, larger \(\alpha_{ij}\) means that species \(i\) and
\(j\) experience more similar environmental fluctuations, which tends to generate
positive abundance correlations. On the other hand, the same larger
\(\alpha_{ij}\) also means stronger competition, which tends to generate negative
abundance correlations. When the noise covariance is chosen to be exactly
\(C^{(\eta)}=\alpha\), these two effects cancel in the linearized dynamics, leaving a
diagonal abundance covariance matrix.

Thus, within this stochastic Lotka-Volterra framework, abundance correlations
contain no direct information about niche overlap under the biologically grounded
prescription \(C^{(\eta)}_{ij}=\alpha_{ij}\). This provides the analytical basis for the
caution that abundance correlations should not be interpreted as direct proxies
for biological niche overlap when the system is modeled using a standard
stochastic Lotka-Volterra framework.

\end{document}